\documentstyle[aps, twocolumn, epsfig, psfig]{revtex}

\draft
\preprint{}
\begin{document}
\twocolumn[\hsize\textwidth\columnwidth\hsize\csname@twocolumnfalse\endcsname
\title{High Pressure Insulator-Metal Transition in Molecular Fluid Oxygen}
\author{Marina Bastea\cite{mb}, Arthur C. Mitchell and William J. Nellis}
\address{Lawrence Livermore National Laboratory, P. O. Box 808, Livermore, CA 94550}
\centerline{\bf Phys.Rev.Lett. 86, 3108 (2001)}
\maketitle

\begin{abstract}
We report the first experimental evidence for a metallic phase in fluid molecular oxygen.
Our electrical conductivity measurements of fluid oxygen under dynamic quasi-isentropic compression 
show that a non-metal/metal transition occurs at 3.4 fold compression, 4500 K and 1.2 Mbar. We discuss the 
main features of the electrical conductivity dependence on density and temperature and give an interpretation 
of the nature of the electrical transport mechanisms in fluid oxygen at these extreme conditions.   
\end{abstract}
]

The transition of condensed matter between electrically conducting and insulating states is a topic of wide  
scientific interest, whose  relevance ranges from superconductivity to collosal magnetoresistance \cite{review}, to 
more recently thermoelectricity \cite{thermoelectrics}. The metal/insulator transition has received renewed attention 
during the past decade in the context of high pressure research \cite{ashcroft}.  Although much progress has been made 
in developing experimental, theoretical and computational tools appropriate for the study of the metal/insulator 
transition at extreme conditions, our present understanding is mostly phenomenological and still incomplete. Given the 
``simple'' nature of their interactions, the homonuclear diatomic molecular species, e.g. hydrogen 
\cite{hydrogen,billPRB,cauble}, oxygen \cite{nicol,ruoff,japanese}, nitrogen \cite{nitrogen} and the halogens 
\cite{halogens}, have been intensively studied using both static and dynamic high pressure techniques. 

We report in this letter the first experimental evidence for a non-metal/metal transition in the molecular fluid phase of
oxygen under high dynamic compression. Oxygen has very rich physics at high pressure. Since dramatic color changes 
were reported in the $100 kbar$ pressure range \cite{nicol1}, the high pressure solid phases of oxygen have been 
extensively investigated using structural \cite{structural,japanese1,italians}, optical \cite{nicol,ruoff} and transport 
techniques \cite{japanese}. A metallic state and evidence for superconductivity have been identified in the solid 
around $1 Mbar$ at very low temperatures \cite{ruoff,japanese}. The properties of the liquid on the other hand have 
not been explored much, due to experimental difficulties and also technical and conceptual challenges for theory. 

New insights on the physics of warm dense matter were provided by dynamic compression experiments on hydrogen 
\cite{billPRB,cauble}. We present the first electrical conductivity measurements of 
fluid oxygen under dynamic quasi-isentropic compression between $0.3$ and $1.9 Mbar$. In our experiments fluid 
oxygen reaches up to 4-fold compression and temperatures below $7000 K$, conditions never before reached 
experimentally. We note that these conditions are similar with the ones found in the interiors of the giant planets, 
where oxygen is a major constituent, and that these conductivity measurements may be instrumental in explaining the 
origins of the planetary magnetic fields. 

We measured the electrical resistance of quasi-isentropically compressed oxygen starting from high purity 
($99.995\%$), disk shaped liquid samples \cite{samplesize} of density $d_0 = 1.202 g/cm^3$ at $T_0 = 77 K$ and 
atmospheric pressure. The high pressures were generated by multiple reflections of a shock wave between two 
single-crystal sapphire anvils which enclose the sample. The initial shock wave was produced by the impact between 
a high velocity projectile onto a stationary target containing the sample. The gradual increase in pressure produced 
by this technique yields quasi-isentropic compression. The experiments were designed to achieve and maintain 
thermodynamic steady state conditions at the final pressure and density for time durations of the order $100-200ns$, 
during which the measurements were taken. We optimized the target geometry and electrodes size and position in order 
to achieve homogeneous samples of macroscopic size. The shock sensors which provide the trigger for the data acquisition 
were placed outside the sample cavity in order to preclude any interference with the measurements. Other details of the 
shock reverberation technique, cryogenic target construction and electrical circuitry were similar to those of Ref. 
\cite{billPRB}. 

\begin{figure}
\centerline{\psfig{file=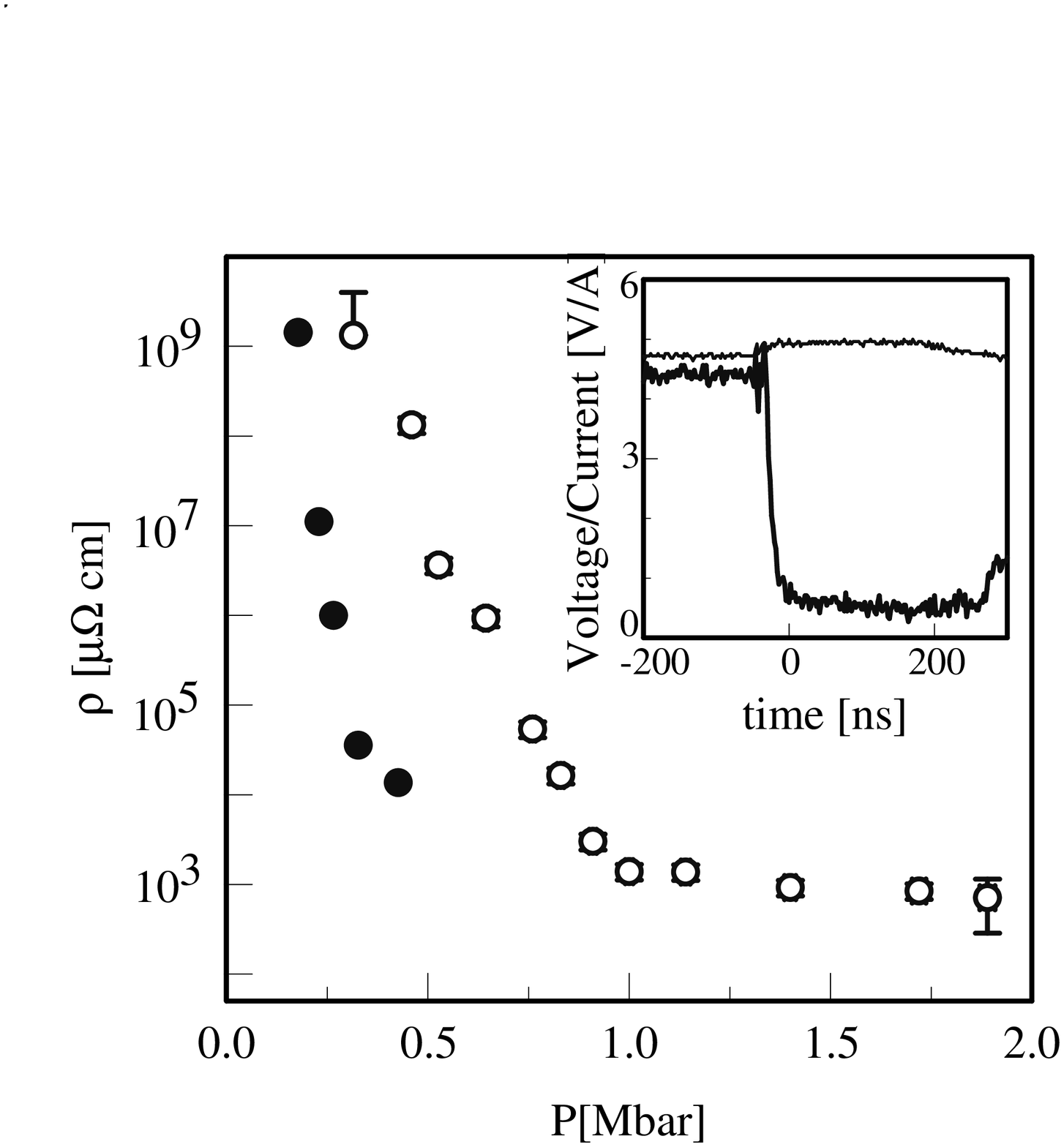,width=2.2truein}}
\caption{Pressure dependence of the electrical conductivity of shock-compressed fluid oxygen; open circles - new 
quasi-isentropic compression data; filled circles - Hugoniot measurements of Hamilton et al.. Error bars are shown where 
larger than symbol size. Inset - example of the measured voltage (lower thick line) and current intensity (upper 
thin line) through the oxygen sample and a parallel $1 \Omega$ shunt resistance.}
\label{ro-p}
\end{figure}

The equilibrium pressure is determined with a $1\%$ accuracy from the measured projectile velocity, using the shock 
impedance matching technique \cite{prescal}. Densities were calculated using a one dimensional hydrodynamic code in 
which the projectile, anvils and sample materials were modeled by Mie-Gruneisen and ratio of polynomials equations of 
state (EOS).
The EOS's were derived based on extensive experimental results of Ref. \cite{lanl,young1}. Temperatures were obtained 
by chemical equilibrium calculations similar with those in \cite{ree}. The sample resistance was extracted from 
accurate measurements of steady voltage and current intensity signals (see Fig.\ref{ro-p} - inset, for an example), 
taken during thermodynamic steady state conditions lasting $\simeq 200 ns$. The electrical resistivity was calculated 
from the sample resistance using $R = C\rho$ where $C$ represents the geometrical factor. The geometrical factor, or cell 
constant, was computed as a finite element solution to the Maxwell equations for the experimental geometry with 
appropriate boundary conditions. The calculations spanned a wide range of sample thicknesses (between $80\mu m$ and 
$200 \mu m$), and were verified by direct measurements performed on materials with known conductivities. 

The measured resistivity of fluid oxygen ranges from $10^{9} \mu \Omega cm$ at $0.3 Mbar$, down to $8 \times 10^{2} 
\mu \Omega cm$ at $1.9 Mbar$. As seen in Fig.\ref{ro-p} these resistivity values are four to five orders of magnitude 
larger than the resistivities observed along the principal shock Hugoniot by Hamilton et al. \cite{hamilton} at the same 
pressure. Such a  big difference can be explained by the fact that the high-pressure states achieved using the shock 
reverberation technique are at significantly lower temperatures ($\sim 1200 K$ at $0.4 Mbar$) than those on the Hugoniot 
($\sim 6500 K$ at $0.4 Mbar$ \cite{hamilton}). In the single shock (Hugoniot) technique, between $0.18 Mbar$ and 
$0.4 Mbar$ the densities increase by $\sim 20 \%$ whereas the temperatures nearly triple. It can be inferred therefore 
that the resistivity measured along the principal Hugoniot is mainly due to a much higher degree of thermal excitation 
of the carriers.  

We distinguish two main regimes for the pressure and density dependence of the electrical resistivity. 
At pressures up to $\sim 1 Mbar$, there is a rapid drop of the resistivity of the compressed material 
which encompasses six orders of magnitude. Between $\sim 1 Mbar$ and $2 Mbar$ the electrical resistivity shows little 
sensitivity to pressure and density variations. In order to understand this behavior we need to analyze the changes 
produced by compression on the electronic configuration and chemical bonding in the warm fluid. The highest occupied 
electronic state in the oxygen molecule at ambient conditions is $\approx 12 eV$ below the continuum. In a crystalline 
solid the energy levels of the isolated molecule split and widen to form sharply separated, alternating bands of allowed 
and prohibited electronic states. In a fluid, due to the inherent structural ``disorder'', the energy 
levels broaden to form bands and the band edges spread into ``tails''. 

\begin{figure}
\centerline{\psfig{file=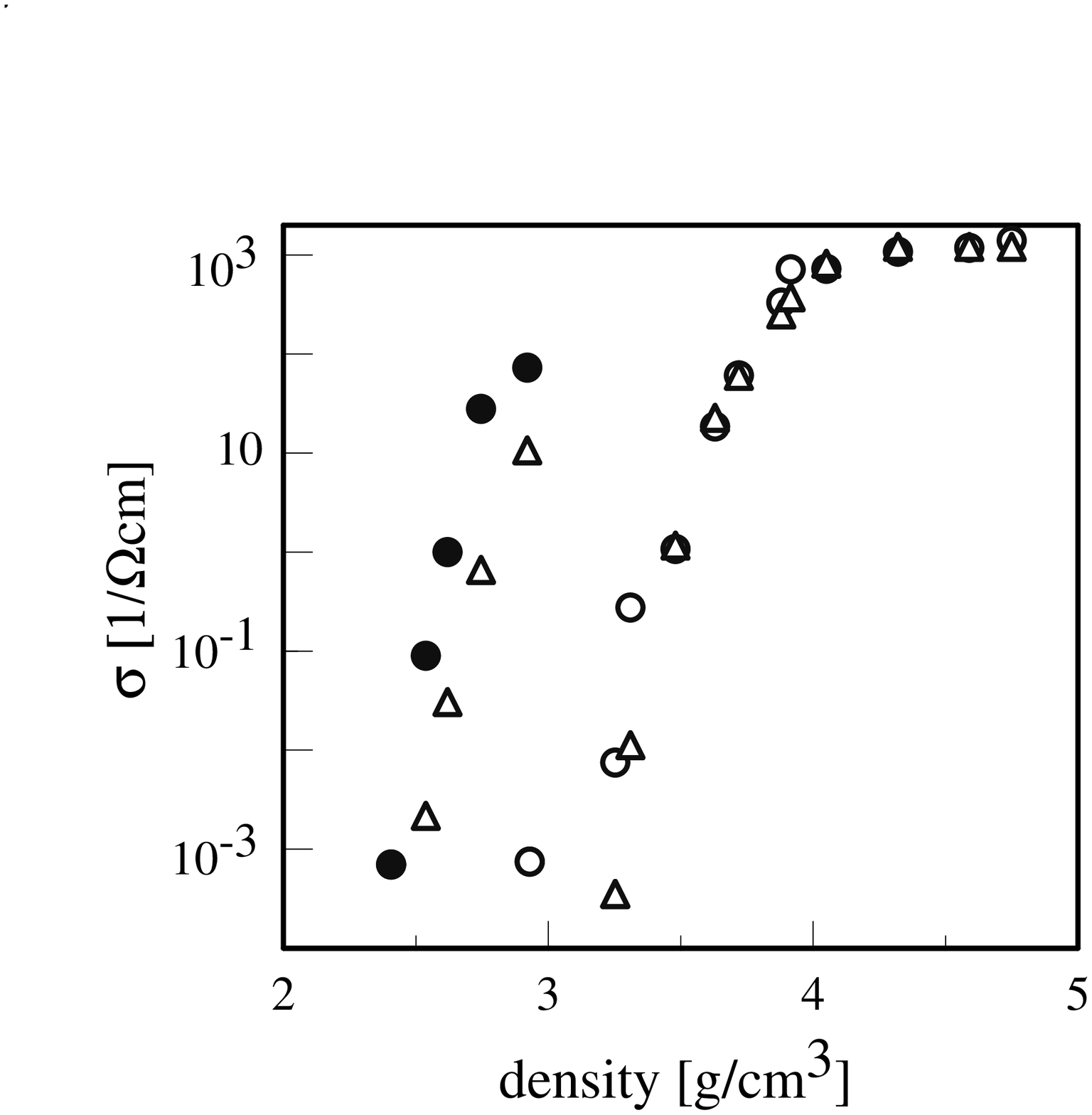,width=2.truein}}
\caption{Density dependence of the electrical conductivity of shock compressed oxygen  - open and filled circles 
for the present shock-reverberation and Hugoniot experiments of Hamilton et al., respectively. Triangles represent 
the model fit (see text)}
\label{sigma-d} 
\end{figure}

\begin{figure}
\centerline{\psfig{file=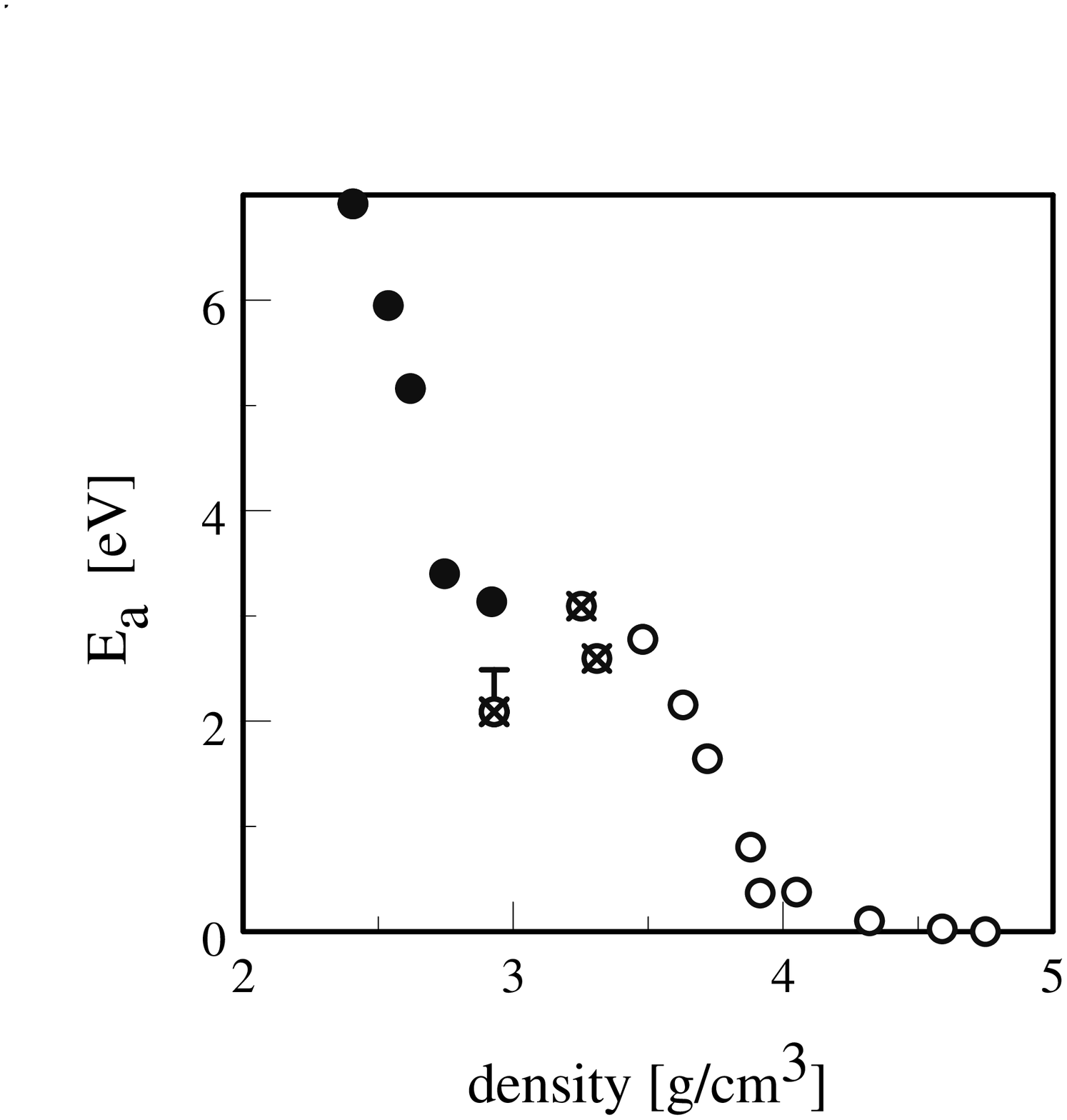,width=2.truein}}
\caption{Activation energy (see text) vs. density - present data (fluid - open circles, amorphous solid - 
crossed circles) and single-shock data of Hamilton et al. (filled circles). Errors bars are smaller than 
symbol size except where shown.} 
\label{Ea-d} 
\end{figure}

Nonetheless, as our experiments also show, liquid 
oxygen at $77 K$ and $1 bar$ is a wide gap electrical insulator. As pressure increases the energy bands 
broaden further and create a non-zero electronic density at the Fermi level. Most of the states in the tails of 
the valence and conduction bands and at the Fermi level are localized, as demonstrated by Mott \cite{mott1}. We
postulate the existence of an activation energy $E_a$ (mobility gap), i.e. the minimum energy that a valence electron 
needs to acquire to move into a  delocalized state and participate in conduction, and use an activation model for the 
electrical conductivity variation with temperature $T$:
\begin{equation}
\sigma = \sigma_o \exp\left[- {{{E}_a(d)}  \over {2 k_B T }} \right]
\end{equation}
\begin{equation}
{E}_a = {E}_o \left[ \left( {\frac{d_m}{d}}\right)^{1/3} - 1\right] {\bf \cal H}(d_m-d)
\label{ea}
\end{equation}
where $\sigma_o$ and $d_m$ are the electrical conductivity and density at the transition into the metallic 
state respectively, and ${\bf \cal H}(d_m-d)$ is the Heaviside step function. 
We assume a linear dependence of the 
activation energy on the distance between neighboring molecules, Eqn. \ref{ea}, and zero value above the 
non-metal/metal transition. This approximation should work best in the vicinity of the transition, as it represents 
the first order expansion of the mobility gap in the deviation from the electronic wave-function overlap at 
metallization. The fit results are shown as triangles in Fig.\ref{sigma-d}. The parameters at which the 
activation energy vanishes are $P_m = 1.2 Mbar$, $d_m = 4.1 g/cm^3$ and $\sigma _o = 1205 /\Omega cm$. Using the 
fitted $\sigma _o$ we extract the activation energy $E_a = 2 k_B T ln(\sigma_o/\sigma)$ from the data, 
Fig.\ref{Ea-d}. Although the activation energy effectively vanishes around $1.2 Mbar$, it becomes in fact comparable 
to the thermal excitation energy at somewhat lower pressures which is consistent with the observed change in slope 
of the experimental data. The conductivity at the transition, $\sigma_o$, is interpreted as the Ioffe-Regel limit 
\cite{ioffe-regel}, corresponding to a scattering length (mean free path) $l$ of the electrons of the order 
of the intermolecular distance. Using the Drude model, we compute the effective electronic density 
$n_{e} = \sigma_o m_e v_e/ e^2 l$, where $v_e = \hbar k/m_e$ is the typical electronic speed, $kl\simeq 1$. We find that the 
effective number of conduction electrons per molecule is $\simeq 0.1$ (for a good metal, e.g. copper at ambient conditions, 
this number is $\simeq 0.5$). 

The fact that the electrical conductivity does not appear to saturate in the metallic state is not unexpected. The 
``resistivity saturation'' concept has been introduced by Fisk and Webb in order to explain the behavior of strongly 
interacting, disordered systems \cite{fiskwebb}. However, there is both experimental and theoretical \cite{millis} 
evidence in support of the absence of saturation. Just as found by Millis et al. in the study of a simple representative 
system of electrons strongly coupled to phonons \cite{millis}, in the case of compressed fluid oxygen there is a smooth 
crossover, as the density is increased, from activated conduction to a transport regime which is dominated by hopping 
between nearest neighbors whose separation distance decreases with further increments of pressure.

In the lower density regime, see Fig.\ref{sigma-d}, there are significant deviations of the model from the measured 
conductivities (see also Fig.\ref{Ea-d}, the activation energy dependence on density). Linear extrapolation of the oxygen 
melting curve from Ref. \cite{young1} indicates that up to approximately $0.55 Mbar$ the states achieved by quasi-isentropic 
compression fall into the solid region of the phase diagram. We believe that, due to the short time-scales of the experiments,
the system does not crystallize but rather reaches a glassy state. The temperature and density calculations that we perform 
using the EOS of the liquid should still describe the system reasonably well. However, as the oxygen molecule is rather 
elongated, the glassy states so obtained may contain a significant degree of orientational order ``frozen in''. We 
speculate that these correlations lead to increased delocalization of the electrons and therefore better conduction. 

The model deviations from the single-shock Hugoniot conductivity measurements are consistent with our expectations 
that as the density is decreased higher order terms in the activation energy ($E_a$) expansion become important. 
Also due to the fact that the Hugoniot states are much hotter, it is expected that the fluid would dissociate and 
contributions to the conductivity would arise from the atomic species as well. Chemical equilibrium calculations similar 
with those in \cite{ree} yield an estimate of the degree of dissociation in the regime of interest. For example, in our 
experiments, at a density of $3.9 g/cm^3$, $1 Mbar$ and $3900 K$ approximately $2.4 \%$ of the molecules break-up. Also, 
at $0.4 Mbar$ and $1200 K$ the shock-reverberated states have a dissociation fraction of less than  $10^{-4}\%$ 
as compared with the $~7\%$ along the principal Hugoniot at the same pressure and $6500 K$. We conclude therefore that the 
fluid undergoing quasi-isentropic compression is mostly molecular. 
We analyze the crossover to the metallic regime from the perspective of the Mott principle, which states that an 
insulator/metal transition may occur when the separation distance between the centers, of number density $n$, that 
produce charge carriers becomes comparable with the spatial extent of the electronic wave functions \cite{mott1}. Upon 
sufficient overlap of the wave functions the outermost electrons become delocalized and participate in conduction. This 
is predicted to occur at a critical number density $n = n_c$, that satisfies $n_c^{1/3} \times a_B^{\star} \simeq 0.25$,
where $a_B^\star$ is a generalized Bohr radius, i.e. the radius where the probability of finding a valence electron is 
maximum. In order to estimate $a_B^\star$ for the $O_2$ molecule we calculated its electronic 
charge density as a function of position using the GAMESS package. We performed restricted open shell Hartree-Fock 
calculations using Gaussian basis sets to determine the electronic structure of the $S=1$ ground state oxygen molecule. 
The $O_2$ molecule is $30\%$  elongated in the direction of the bond. Since in the warm fluid the angular correlations 
between molecules should be relatively small, we calculated an effective electronic charge density as an average of 
$r^2\psi\psi^{\star}$ over the solid angle. We found that for oxygen $<r^2\psi\psi^{\star}>$ as a function of distance 
has a maximum at $1.2 bohr$, which was taken to be the generalized Bohr radius, $a_B^{\star}$.

\begin{figure}
\centerline{\psfig{file=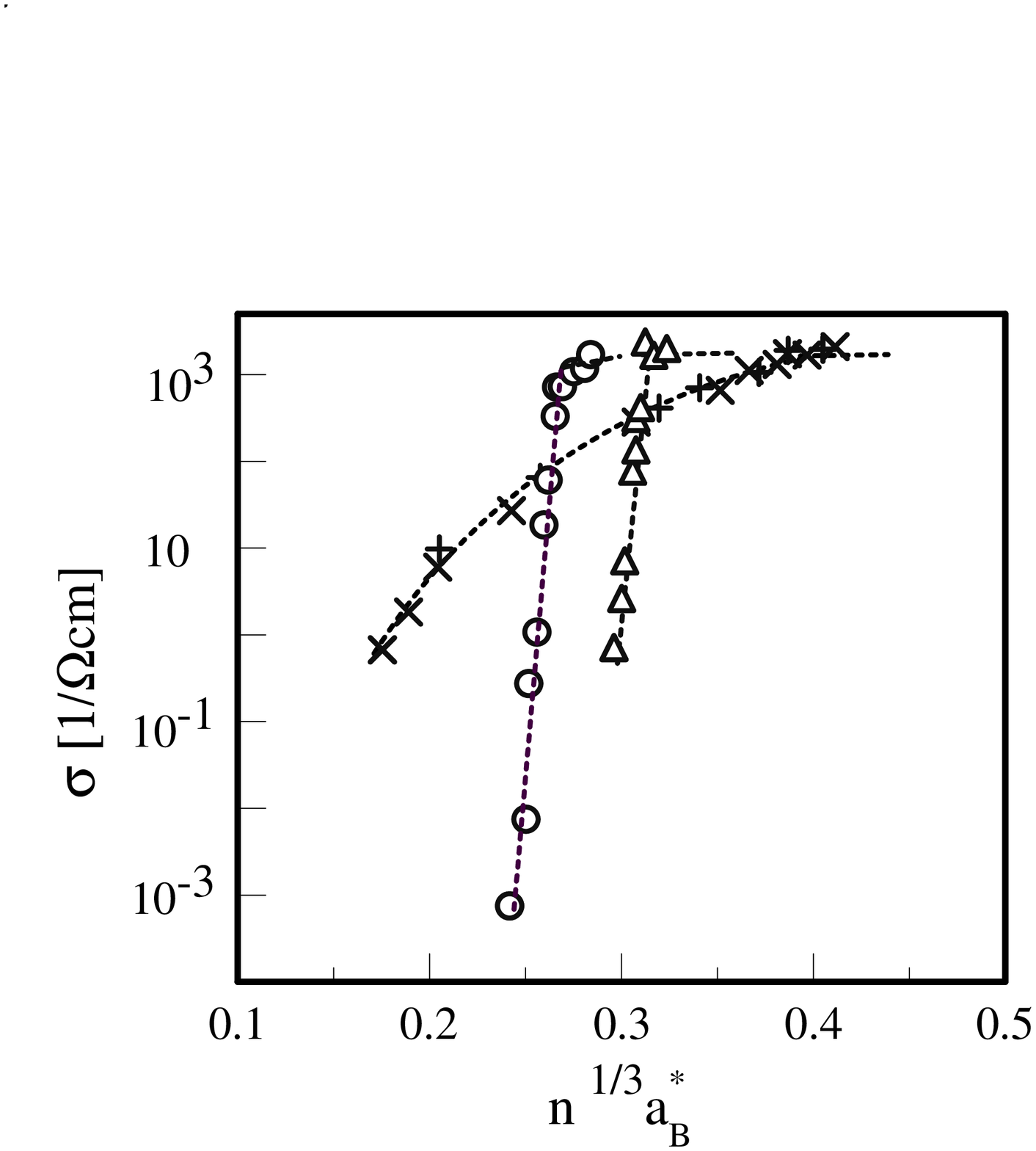,width=2.2truein}}
\caption{Electrical conductivity dependence on the Mott scaling parameter for oxygen (circles), hydrogen (triangles), 
rubidium (crosses) and cesium (x). Dotted lines are guides to the eye.}
\label{mott}
\end{figure}

The electrical conductivity dependence of liquid oxygen on the Mott scaling parameter $n^{1/3}\times a_B^{\star}$ 
is shown in Fig.\ref{mott}. Also shown are the results for hydrogen, cesium and rubidium, which are believed to 
undergo insulator/metal Mott transitions at high pressures \cite{billPRB,edwards}. It is remarkable that for the 
two molecular fluids as well as the two alkalis there are obvious changes in the conductivity at values of the 
Mott scaling parameter which are close to the predicted value for the Mott transition. The slower 
increase of the conductivity in the case of the alkali metals, which have only one relatively weakly bound 
valence electron, can be attributed to the spatially extended nature of their electronic wave functions. 

The transition of fluid oxygen to a metallic state should be accompanied by the divergence of the dielectric 
susceptibility $\epsilon$, also known as the Goldhammer-Herzfeld criterion \cite{GoldHerz}. Given the limitations of the 
Clausius-Mossoti equation \cite{martin}, $\epsilon=(1+8\pi n \alpha/3)/(1-4\pi n \alpha/3)$ 
($n$ is $O_2$ number density and $\alpha$ is the molecular polarizability), we use it only to gain a 
qualitative understanding. We find that at metalization, at a density $d = 4.1 g/cm^3$, $\alpha\approx 3.17 \AA ^3$, a 
factor of two bigger than the measured polarizability of molecular oxygen at standard ambient conditions \cite{newell}. 
A similar effect has been observed in diatomic liquid mercury \cite{edwards}. We interpret the enhancement of the $O_2$
molecular polarizability as an indication that the molecules at these conditions are distorted.

Although, to our knowledge, there are no theoretical studies of fluid oxygen in the Mbar pressure range, several 
predictions have been made \cite{italians} for the solid in conjunction with the reported metallization of solid 
oxygen under static high pressure \cite{ruoff,japanese}. The reported metallization pressure for the low temperature 
solid, $0.96 Mbar$, is very close to the onset of the metallic conduction in the fluid around $1.2 Mbar$. However, the 
mechanisms that lead to the non-metal/metal transition in the fluid are fundamentally different from the solid. In the 
fluid we observe a Mott transition where density driven band closure and disorder play the fundamental roles. In the 
solid, on the other hand, the metallization is associated with a structural phase transition \cite{japanese1}.

We would like to thank N. Ashcroft, M. Nicol, G. Galli, S. Bastea and A.J. Millis for useful discussions, 
D. Young for the oxygen EOS table, J. Reaugh for the hydrocode and N. Winter for help with the GAMESS 
code. We gratefully acknowledge the LDRD office for financial support. We also thank S. Caldwell, W.P. Hall, N.Hisey, 
K. Stickle, L. Raper and T. Uphaus for assistance at the gas-gun facility. This work was performed under the auspices 
of the U. S. Department of Energy by University of California Lawrence Livermore National Laboratory under Contract 
No. W-7405-Eng-48.


\begin{references}
\bibitem{samplesize}The samples have a diameter of $25.4 mm$ and initial thicknesses of $0.32 mm$ to 
$0.38 mm$.
\end{references}
\end{document}